\documentclass[cpp,a4paper,fleqn]{w-art}

\usepackage{times,cite,w-thm}
\usepackage[]{graphicx}

\begin{document}

\DOIsuffix{theDOIsuffix}

\Volume{}
\Month{}
\Year{}

\pagespan{1}{}

\Receiveddate{XXXX}
\Reviseddate{XXXX}
\Accepteddate{XXXX}
\Dateposted{XXXX}


\title[Long-range correlations of the surface charge density between 
electrical media]{Long-Range Correlations of the Surface Charge Density 
Between Electrical Media with Flat and Spherical Interfaces}

\author[L. \v{S}amaj]{Ladislav \v{S}amaj\footnote{Invited talk at 
the conference SCCS 2011, Budapest; E-mail:~\textsf{Ladislav.Samaj@savba.sk}}}
\address[]{Institute of Physics, Slovak Academy of Sciences, D\'ubravsk\'a
cesta 9, 845 11 Bratislava, Slovakia}

\begin{abstract}
We study the asymptotic long-range behavior of the time-dependent correlation 
function of the surface charge density induced on the interface between 
two media of distinct dielectric functions which are in thermal equilibrium
with one another as well as with the radiated electromagnetic field. 
We start with a short review which summarizes the results obtained by using 
quantum and classical descriptions of media, in both non-retarded and 
retarded regimes of particle interactions. 
The classical static result for the flat interface is rederived by using a
combination of the microscopic linear response theory and the macroscopic
method of electrostatic image charges.
The method is then applied to the case of a spherically shaped interface
between media.
\end{abstract}

\maketitle

\section{A short review} \label{Introduction}
This article is about a simple inhomogeneous physical system consisting of 
two distinct electrical media in contact with one another which are in thermal
equilibrium.
Such system is easily accessible in experiments and can serve as a tool 
for testing quantum mechanics (and its classical limit) and crossover 
between retardation and non-retarded effects.

\begin{figure}[h]
\begin{minipage}{72mm}
\includegraphics[width=.72\textwidth,height=30mm]{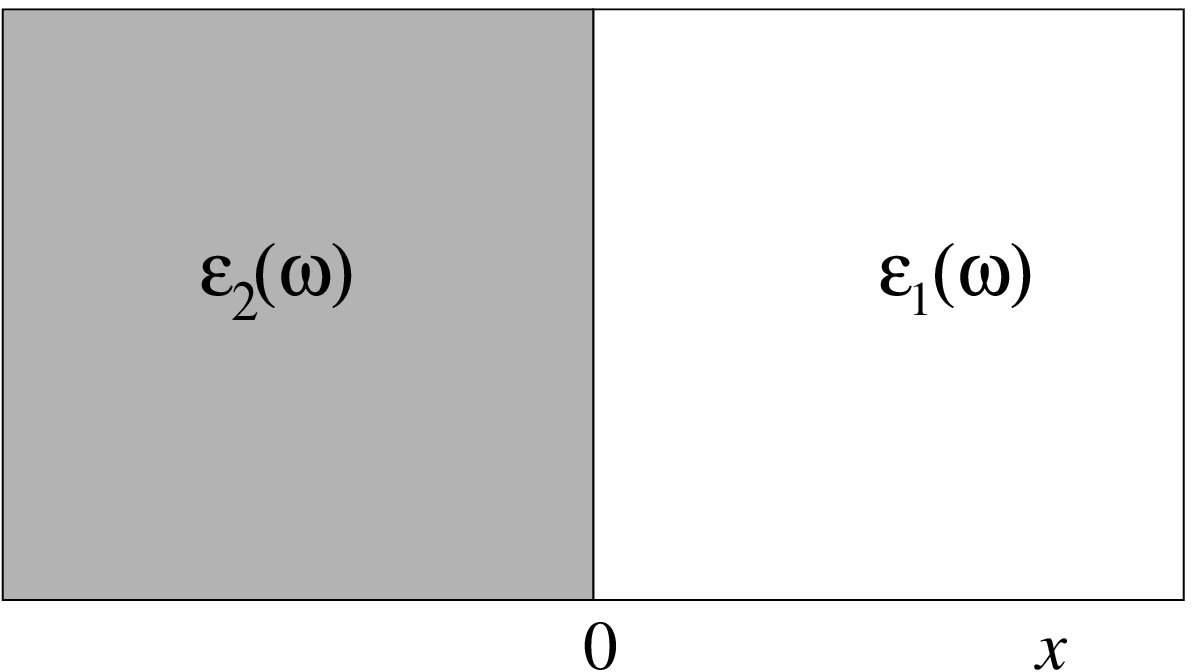}
\caption{The geometry of two semi-infinite electrical media with the flat 
interface contact at $x=0$.}
\label{fig:1}
\end{minipage}
\hfil
\begin{minipage}{72mm}
\includegraphics[width=.44\textwidth,height=30mm]{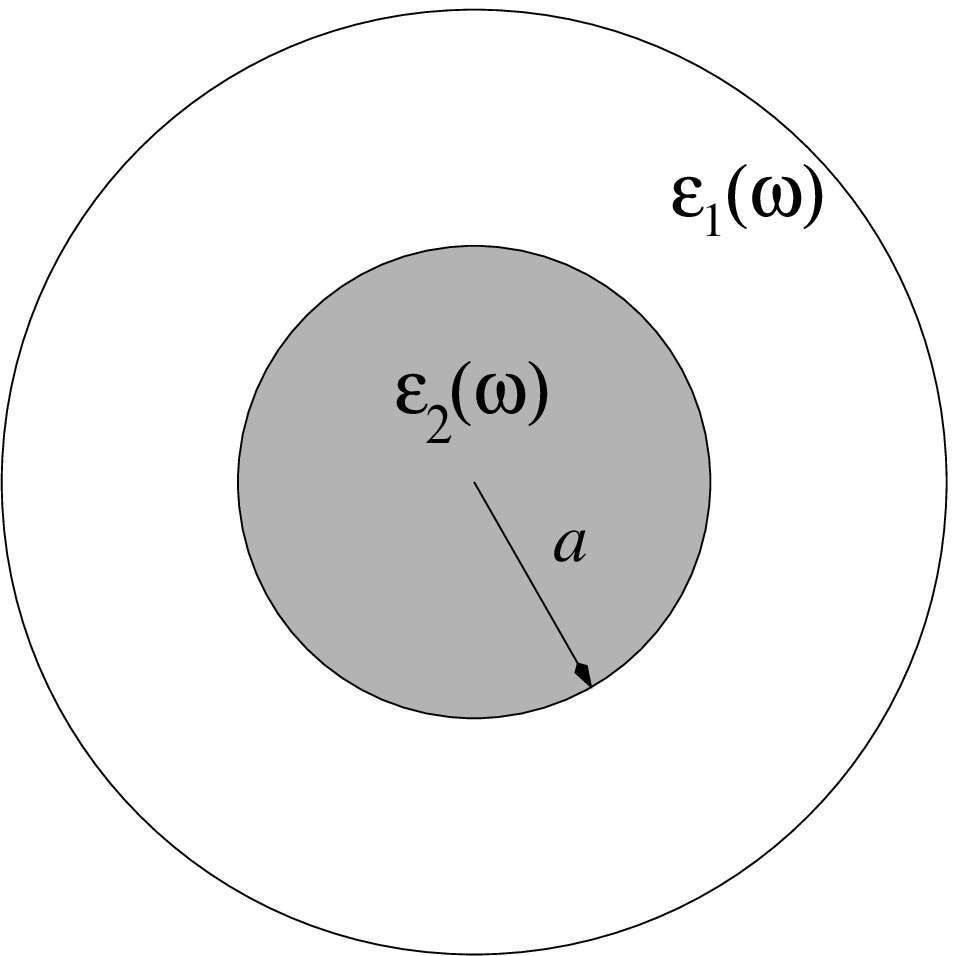}
\caption{The geometry of two electrical media with the spherical interface 
of radius $a$.}
\label{fig:2}
\end{minipage}
\end{figure}

The special geometry of two semi-infinite media with a flat interface, 
formulated in the three-dimensional Cartesian space of points 
${\bf r}=(x,y,z)$, is pictured in Fig. \ref{fig:1}.
The two media with the distinct frequency-dependent dielectric functions
$\epsilon_1(\omega)$ and $\epsilon_2(\omega)$ occupy
the complementary half-spaces $x>0$ and $x<0$, respectively.
The flat interface between media is the plane $x=0$.
We recall that, in Gauss units, $\epsilon(\omega)=1$ for vacuum, 
while the static dielectric constant $\epsilon(0)$ is finite ($>1$)
for dielectrics and diverging, $\epsilon(0)\to {\rm i}\infty$, for conductors.
The system is translationally invariant along each plane formed by
the two coordinates ${\bf R}=(y,z)$ perpendicular to $x$.
We assume that the media have no magnetic structure and the magnetic
permeabilities $\mu_1=\mu_2=1$.
Non-relativistic charged particles forming the two media and the radiated 
electromagnetic (EM) field are in thermal equilibrium at some temperature $T$, 
or the inverse temperature $\beta=1/(k_{\rm B}T)$ with $k_{\rm B}$ 
being the Boltzmann constant. 

The different electric properties of the media give rise to a microscopic 
surface charge density (operator in quantum mechanics) $\sigma(t,{\bf R})$ 
at time $t$ and point ${\bf r}=(0,{\bf R})$ on the interface.
It is understood as being the microscopic volume charge density integrated
along the $x$-axis on some microscopic depth from the interface.
According to the elementary electrodynamics, the surface charge density
is associated with the discontinuity of the normal $x$ component of
the microscopic electric field at the interface:
\begin{equation} \label{surfacecharge}
\sigma(t,{\bf R}) = \frac{1}{4\pi} \left[
E_x^+(t,{\bf R}) - E_x^-(t,{\bf R}) \right] ,
\end{equation}
where the superscript $+/-$ means approaching the interface through
the limit $x\to 0^+/0^-$.
The tangential $y$ and $z$ components of the electric field are continuous
at the interface.
The correlation of fluctuations of the surface charge density at two
points on the interface, with times different by $t$ and distances
different by $R=\vert {\bf R}\vert$, is described by the 
(symmetrized) correlation function
\begin{equation} \label{structurefunction}
S(t,R) \equiv \frac{1}{2} \left\langle
\sigma(t,{\bf R}) \sigma(0,{\bf 0}) 
+ \sigma(0,{\bf 0}) \sigma(t,{\bf R}) \right\rangle^{\rm T} ,
\end{equation}
where $\langle\cdots\rangle^{\rm T}$ represents a truncated equilibrium
average at the inverse temperature $\beta$,
$\langle A B \rangle^{\rm T} = \langle A B \rangle -
\langle A \rangle \langle B \rangle$.
With regard to (\ref{surfacecharge}), this correlation function is related to 
fluctuations of the electric field on the interface.

Although the system is not in a critical state, the combination of
the spatial inhomogeneity and long-ranged EM interactions in the media 
causes that the asymptotic large-distance behavior of the surface charge 
correlation function (\ref{structurefunction}) exhibits, in general,
the long-range tail of type
\begin{equation} \label{asymptotic}
\beta S(t,R) \sim \frac{h(t)}{R^3} , \qquad R\to\infty
\end{equation}
with some prefactor (slope) function $h(t)$.
It is useful to introduce the two-dimensional Fourier transform
$S(t,q) = \int {\rm d}^2R \exp(-{\rm i}{\bf q}\cdot{\bf R}) S(t,{\bf R})$ ,
where the wave vector ${\bf q} = (q_y,q_z)$.
In the sense of distributions, the Fourier transform of $1/R^3$ is $-2\pi q$.
Consequently, $\beta S(t,q)$ has a kink singularity at ${\bf q}={\bf 0}$:
$\beta S(t,q) \sim -2\pi h(t) q$ for $q\to 0$.
$S(t,q)$ is measurable by scattering experiments.
The fact that it is linear in $q$ for small $q$ makes this quantity very
different from the usual bulk structure functions with short-range 
(usually exponential) decays which are proportional to $q^2$ in 
the limit $q\to 0$.   

The form of the prefactor function $h(t)$ depends on ``the physical model'' 
used.
Defining media, one can apply:
\begin{itemize}
\item
Either {\em classical} mechanics, $h_{\rm cl}(t)$, or {\em quantum} mechanics,
$h_{\rm qu}(t)$.
According to the correspondence principle, the quantum mechanics
reduces to the classical one in the high-temperature limit.
\item
Either {\em non-retarded} description, $h^{(\rm nr)}$, or {\em retarded} 
description, $h^{(\rm r)}$, of particle interactions.
In the non-retarded regime, the speed of light is taken to be infinite,
$c=\infty$, ignoring in this way magnetic forces.
In the retarded regime, $c$ is assumed finite and so the charged particles
are fully coupled to both electric and magnetic parts of the radiated EM
field.
According to the Bohr-Van Leeuwen theorem \cite{BVtheorem}, magnetic
degrees of freedom can be effectively eliminated from statistical
averages of classical systems for the static case $t=0$, so that
$h_{\rm cl}^{(\rm r)}(0) = h_{\rm cl}^{(\rm nr)}(0) \equiv h_{\rm cl}(0)$.   
If the EM field is thermalized, it was shown in Ref. \cite{Alastuey00}
that the decoupling arises only at distances larger than the thermal
photon wavelength $\beta\hbar c$, where the EM field itself can be
treated classically.
Quantum electrodynamics must be used at shorter distances, which
invalidates the decoupling.
At room temperature, the wavelength $\beta\hbar c$ is rather large
compared to usual microscopic scales and might have some experimental
importance.
The decoupling is preserved in our macroscopic electrodynamics.
For time differences $t\ne 0$, magnetic interactions do affect classical 
charged systems in thermal equilibrium and so, in general,
$h_{\rm cl}^{(\rm r)}(t) \ne h_{\rm cl}^{(\rm nr)}(t)$.   
\end{itemize}

The media configuration studied in the past was exclusively a conductor 
in contact with vacuum.
When the conductor is modelled by a {\em classical} Coulomb fluid,
a microscopic analysis \cite{Jancovici82} leads to the universal static result 
\begin{equation} \label{universal}
h_{\rm cl}(0) = - \frac{1}{8\pi^2} \qquad \mbox{(conductor vs. vacuum),}
\end{equation}
independent of the fluid composition.
The same result has been obtained later \cite{Jancovici95} by using simple
macroscopic arguments. 
As concerns {\em quantum} microscopic models of Coulomb fluids,
a substantial simplification arises in the so-called jellium
(one-component plasma), i.e. the system of identical pointlike particles
of charge $e$, mass $m$ and bulk number density $n$, immersed in a
fixed uniform neutralizing background of charge density $-e n$.
The dynamical properties of the jellium have a special feature: There is no 
viscous damping of the long-wavelength plasma oscillations \cite{Pines66}.
In the non-retarded regime of the Maxwell equations, the frequencies of 
nondispersive long-wavelength collective modes, namely $\omega_p$ of the bulk 
plasmons and $\omega_s$ of the surface plasmons, are given by
$\omega_p = \sqrt{4\pi n e^2/m}$ and $\omega_s = \omega_p/\sqrt{2}$.
The dielectric function of the jellium is described by a one-resonance
Drude formula
\begin{equation} \label{jellium}
\epsilon(\omega) = 1 - \frac{\omega_p^2}{\omega(\omega+{\rm i}\eta)} ,
\end{equation}
where the dissipation constant $\eta$ is taken as positive infinitesimal,
$\eta\to 0^+$.
The obtained time-dependent result has the non-universal form
\cite{JancoviciLebowitz}
\begin{equation} \label{nonuniversal}
h_{\rm qu}^{(\rm nr)}(t) = - \frac{1}{8\pi^2} \left[ 
2 g(\omega_s) \cos(\omega_s t) - g(\omega_p) \cos(\omega_p t) \right] 
\qquad \mbox{(jellium vs. vacuum),}
\end{equation}
where the function
\begin{equation}
g(\omega) \equiv \frac{\beta\hbar\omega}{2} 
\coth\left( \frac{\beta\hbar\omega}{2} \right) 
= 1 + \sum_{j=1}^{\infty} \frac{2\omega^2}{\omega^2 + \xi_j^2} ,
\qquad \xi_j = \frac{2\pi}{\beta\hbar} j .
\end{equation}
In the complex upper half-plane, $g(\omega)$ possesses an infinite sequence 
of simple poles at the (imaginary) Matsubara frequencies 
${\rm i}\xi_j$ $(j=1,2,\ldots)$.
In the classical limit $\beta\hbar\omega\to 0$, the function $g(\omega)\to 1$ 
for any $\omega$ and the quantum formula (\ref{nonuniversal}) reduces to
\begin{equation} \label{quantum}
h_{\rm cl}^{(\rm nr)}(t) = - \frac{1}{8\pi^2} \left[ 
2 \cos(\omega_s t) - \cos(\omega_p t) \right] .
\end{equation}
For $t=0$, we recover the universal classical static result (\ref{universal}). 

To study a general configuration of two media in contact and the effect
of retardation, in a series of works \cite{Samaj08,Jancovici09} we applied 
the macroscopic fluctuational electrodynamics of Rytov \cite{Rytov} 
to the present problem. 
The classical static result was found in the form
\begin{equation} \label{classstat}
h_{\rm cl}(0) = - \frac{1}{8\pi^2} \left( \frac{1}{\epsilon_1}
+ \frac{1}{\epsilon_2} - \frac{4}{\epsilon_1+\epsilon_2} \right) ,
\end{equation}
where $\epsilon_1\equiv \epsilon_1(0)$ and $\epsilon_2\equiv \epsilon_2(0)$.
For the previously studied configuration of a conductor 
$(\epsilon_1\to {\rm i}\infty$ and vacuum $(\epsilon_2=1)$, 
we recover the universal formula (\ref{universal}).
In the retarded regime, for any $t$ we found that 
\begin{equation} \label{quantumr}
h_{\rm qu}^{(\rm r)}(t) = h_{\rm cl}(0),
\end{equation} 
independently of $\hbar$, $c$ and the temperature.
This surprising result was verified explicitly on a microscopic model of 
jellium with retarded interactions \cite{Jancovici09}.
In the non-retarded regime, we got
\begin{equation} \label{non-retard}
h_{\rm qu}^{(\rm nr)}(t) = - \frac{1}{8\pi^2} \left[ \frac{1}{\epsilon_1}
+ \frac{1}{\epsilon_2} - \frac{4}{\epsilon_1+\epsilon_2} 
+ {\rm Re}\, f({\rm i}t) \right] ,
\end{equation}
where $f({\rm i}t)$ is an analytic $\tau\to {\rm i}t$ continuation
of the function
\begin{equation} \label{fction}
f(\tau) = 2 \sum_{j=1}^{\infty} \left[ \frac{1}{\epsilon_1({\rm i}\xi_j)}
+ \frac{1}{\epsilon_2({\rm i}\xi_j)} - 
\frac{4}{\epsilon_1({\rm i}\xi_j)+\epsilon_2({\rm i}\xi_j)} \right]
\cos\left( \xi_j\tau \right) .
\end{equation}
Plugging into this expression the dielectric function of the jellium
(\ref{jellium}), one recovers the microscopic result (\ref{nonuniversal}).
A crossover between the retarded and non-retarded regions was found
at $q_{\rm cross}\sim \omega_p/c$: If $q<q_{\rm cross}$ $(q>q_{\rm cross})$,
the slope function $h^{(\rm r)}$ $(h^{(\rm nr)})$ takes place.

After the short overview, we shall investigate how the classical static
result (\ref{classstat}) modifies itself for a spherically curved interface 
between two media (see Fig. \ref{fig:2}).
Firstly, in section 2, we provide an alternative derivation of
(\ref{classstat}) for the flat interface by using a combination of 
microscopic and macroscopic approaches.
This method allows us to generalize the classical static result
to the spherical interface between media in section 3.

\section{Rederivation of the classical static result (\ref{classstat}) 
for the flat interface} \label{flat}
Our strategy is to compute, for an arbitrary configuration of two points 
${\bf r}$ and ${\bf r}'$ in the media 1 and 2, the correlation function
$\langle \phi({\bf r}) \phi({\bf r}') \rangle$, where $\phi({\bf r})$
is the microscopic electric potential created by constituents of the media
at point ${\bf r}$.
It is related to the microscopic charge density $\rho$ by
$\phi({\bf r}) = \int {\rm d}{\bf r}'' 
\rho({\bf r}'')/\vert {\bf r}-{\bf r}'' \vert$,
where the integral is over the whole space.
Since ${\bf E}({\bf r}) = - \nabla \phi({\bf r})$, we have
\begin{equation} \label{field}
\langle E_x({\bf r}) E_x({\bf r}') \rangle^{\rm T} =
\frac{\partial^2}{\partial x \partial x'}
\langle \phi({\bf r}) \phi({\bf r}') \rangle^{\rm T} .
\end{equation}
According to the relation (\ref{surfacecharge}), the surface charge
correlation function is given by
\begin{equation} \label{surfacechargecorr}
\langle \sigma({\bf R}) \sigma({\bf R}') \rangle^{\rm T} = \frac{1}{(4\pi)^2}
\langle E_x^+({\bf R}) E_x^+({\bf R}') + E_x^-({\bf R}) E_x^-({\bf R}')
- E_x^+({\bf R}) E_x^-({\bf R}') - E_x^-({\bf R}) E_x^+({\bf R}') 
\rangle^{\rm T} .
\end{equation}

Let us introduce an infinitesimal test charge $q$ at point ${\bf r}$.
Microscopically, denoting by $\phi_{\rm tot}({\bf r}')$ the {\em total} 
(i.e. direct plus due to the particles) microscopic potential induced at 
point ${\bf r}'$, it holds 
$\langle \phi_{\rm tot}({\bf r}') \rangle_q = 
q/\vert {\bf r}-{\bf r}' \vert + \langle \phi({\bf r}') \rangle_q$,
where $\langle \cdots \rangle_q$ means an equilibrium average 
in the presence of charge $q$.
The additional Hamiltonian is $\delta H = q\phi({\bf r})$.
Using the classical linear response theory for 
$\langle \phi({\bf r}') \rangle_q$, we obtain
\begin{equation} \label{linearresponse}
\langle \phi({\bf r}') \rangle_q = \langle \phi({\bf r}') \rangle 
- \beta \langle \phi({\bf r}') \delta H \rangle^{\rm T} = 
\langle \phi({\bf r}') \rangle
- \beta q \langle \phi({\bf r}') \phi({\bf r}) \rangle^{\rm T} ,
\end{equation}
where $\langle \cdots \rangle$ means the standard equilibrium average in 
the absence of the test charge $q$. 
Consequently,
\begin{equation} \label{shift}
\beta \langle \phi({\bf r}) \phi({\bf r}') \rangle^{\rm T} =
\frac{1}{\vert {\bf r}'-{\bf r} \vert} 
- \frac{1}{q} \left[ \langle \phi_{\rm tot}({\bf r}') \rangle_q 
- \langle \phi({\bf r}') \rangle \right] . 
\end{equation}

The shift of the mean potential at ${\bf r}'$ due to the test charge $q$ at 
${\bf r}$,
$\langle \phi_{\rm tot}({\bf r}') \rangle_q - \langle \phi({\bf r}') \rangle$,
is determined by using the macroscopic method of images 
\cite{Jackson,Schwinger}.
If both points ${\bf r}$ and ${\bf r}'$ are inside medium 1, the shift 
is given by
\begin{equation} \label{image11}
\langle \phi_{\rm tot}({\bf r}') \rangle_q - \langle \phi({\bf r}') \rangle
= \frac{q}{\epsilon_1\vert {\bf r}'-{\bf r}\vert} +
\frac{q'}{\epsilon_1\vert {\bf r}'-{\bf r}^*\vert} , \qquad
q' = \frac{\epsilon_1-\epsilon_2}{\epsilon_1+\epsilon_2} q ,
\end{equation}
where ${\bf r}^* = (-x,{\bf R})$ is the position of the image charge $q'$.
If both points ${\bf r}$ and ${\bf r}'$ are inside medium 2,
the formula (\ref{image11}) with interchanged indices 1 and 2 applies.
If ${\bf r}$ is in medium 1 and ${\bf r}'$ is in medium 2, or vice versa,
it holds
\begin{equation} \label{image12}
\langle \phi_{\rm tot}({\bf r}') \rangle_q - \langle \phi({\bf r}') \rangle
= \frac{q''}{\epsilon_2\vert {\bf r}'-{\bf r}\vert} , \qquad
q'' = \frac{2\epsilon_2}{\epsilon_1+\epsilon_2} q .
\end{equation}

Inserting the shift of the mean potential to (\ref{shift}) for all four 
possible configurations of points ${\bf r}$ and ${\bf r}'$ in media 1 and 2, 
using (\ref{field}) and the relations
\begin{equation}
\frac{\partial^2}{\partial x\partial x'} 
\frac{1}{\vert {\bf r}'-{\bf r}\vert}\Big\vert_{x=x'=0}
= \frac{1}{\vert {\bf R}-{\bf R}'\vert^3} , \qquad
\frac{\partial^2}{\partial x\partial x'} 
\frac{1}{\vert {\bf r}'-{\bf r}^*\vert}\Big\vert_{x=x'=0}
= - \frac{1}{\vert {\bf R}-{\bf R}'\vert^3} ,
\end{equation}  
we find that
\begin{equation}
\beta \langle E_x^{\pm}({\bf R}) E_x^{\pm}({\bf R}') \rangle^{\rm T}
= \left( 1 - \frac{2}{\epsilon_{1,2}} + \frac{2}{\epsilon_1+\epsilon_2} \right)
\frac{1}{\vert {\bf R}-{\bf R}'\vert^3}  
\end{equation}
\begin{equation}
\beta \langle E_x^{\pm}({\bf R}) E_x^{\mp}({\bf R}') \rangle^{\rm T}
= \left( 1 - \frac{2}{\epsilon_1+\epsilon_2} \right)
\frac{1}{\vert {\bf R}-{\bf R}'\vert^3} .
\end{equation}
Considering these correlators in (\ref{surfacechargecorr}), we end up
with the classical static result (\ref{classstat}). 

\section{Classical static result for a spherical interface between 
media} \label{curved}
The same method is now applied to the spherical interface of radius $a$ 
between media 1 $(r>a)$ and 2 $(r<a)$, see Fig. \ref{fig:2}.
The microscopic formula (\ref{shift}) remains unchanged.
The macroscopic shift of the mean potential at ${\bf r}'$ due to the test 
charge $q$ at ${\bf r}$ for the spherical geometry can be expanded in terms of 
the Legendre polynomials $\{ P_l(\cos\theta) \}_{l=0}^{\infty}$, with 
$\theta$ being the angle between ${\bf r}$ and ${\bf r}'$, as \cite{Schwinger}
\begin{equation}
\frac{1}{q} \left[ \langle \phi_{\rm tot}({\bf r}') \rangle_q 
- \langle \phi({\bf r}') \rangle \right] = 
\sum_{l=0}^{\infty} g_l(r,r') P_l(\cos\theta) .   
\end{equation}
Denoting by $r_<$ $(r_>)$ the smaller (larger) of $r=\vert{\bf r}\vert$ and
$r'=\vert{\bf r}'\vert$, one has
\begin{equation}
g_l(r,r') = \frac{1}{\epsilon_1} \left[ r_<^l +
\frac{(\epsilon_1-\epsilon_2)l}{\epsilon_1(l+1)+\epsilon_2l}
\frac{a^{2l+1}}{r_<^{l+1}} \right] \frac{1}{r_>^{l+1}} \qquad
\mbox{for $r>a$ and $r'>a$,}
\end{equation}
\begin{equation}
g_l(r,r') = \frac{1}{\epsilon_2} r_<^l \left[ \frac{1}{r_>^{l+1}} +
\frac{(\epsilon_2-\epsilon_1)(l+1)}{\epsilon_1(l+1)+\epsilon_2l}
\frac{r_>^l}{a^{2l+1}} \right] \qquad \mbox{for $r<a$ and $r'<a$,}
\end{equation}
\begin{equation}
g_l(r,r') = \frac{2l+1}{\epsilon_1(l+1)+\epsilon_2l} \frac{{r'}^l}{r^{l+1}}
\qquad \mbox{for $r>a$ and $r'<a$.}
\end{equation}
We define the image position of a charge at ${\bf r}$ as ${\bf r}^* 
= (a/r)^2 {\bf r}$ and consider the summation formulas \cite{Gradshteyn}
\begin{equation}
\frac{1}{\vert {\bf r}-{\bf r}'\vert} 
= \frac{1}{\sqrt{r^2+{r'}^2-2rr'\cos\theta}}
= \sum_{l=0}^{\infty} P_l(\cos\theta) \frac{r_<^l}{r_>^{l+1}} ,
\end{equation}
\begin{equation}
\frac{a}{r} \frac{1}{\vert {\bf r}^*-{\bf r}'\vert} 
= \frac{1}{\sqrt{a^2+(rr')^2/a^2-2rr'\cos\theta}}
= \left\{
\begin{array}{ll}
\sum_{l=0}^{\infty} P_l(\cos\theta) \frac{a^{2l+1}}{(rr')^{l+1}} &
\mbox{if $r>a$ and $r'>a$,} \cr
\sum_{l=0}^{\infty} P_l(\cos\theta) \frac{(rr')^l}{a^{2l+1}} &
\mbox{if $r<a$ and $r'<a$,} 
\end{array} \right. 
\end{equation}
\begin{equation}
\sum_{l=0}^{\infty} P_l(\cos\theta) \frac{t^l}{l+\alpha}
= \int_0^1 {\rm d}x \frac{x^{\alpha-1}}{\sqrt{1+(tx)^2-2tx\cos\theta}}
\qquad \mbox{for $\vert t\vert<1$ and $\alpha>0$,}
\end{equation}
to obtain, using (\ref{shift}), the two-point electric potential 
correlation functions:
\begin{eqnarray}
\beta \langle \phi({\bf r}) \phi({\bf r}') \rangle^{\rm T}
& = & \left( 1-\frac{1}{\epsilon_1} \right) 
\frac{1}{\vert {\bf r}-{\bf r}'\vert} 
+ \frac{\epsilon_2-\epsilon_1}{\epsilon_1(\epsilon_1+\epsilon_2)}
\frac{a}{r} \frac{1}{\vert {\bf r}^*-{\bf r}'\vert} \nonumber\\
& & + \frac{\epsilon_1-\epsilon_2}{(\epsilon_1+\epsilon_2)^2}
\int_0^1 {\rm d}x \frac{x^{-\nu}}{\sqrt{a^2x^2+(rr')^2/a^2-2rr'x\cos\theta}}
\end{eqnarray}
for $r>a$ and $r'>a$,
\begin{eqnarray}
\beta \langle \phi({\bf r}) \phi({\bf r}') \rangle^{\rm T}
& = & \left( 1-\frac{1}{\epsilon_2} \right) 
\frac{1}{\vert {\bf r}-{\bf r}'\vert} 
+ \frac{\epsilon_1-\epsilon_2}{\epsilon_2(\epsilon_1+\epsilon_2)}
\frac{a}{r} \frac{1}{\vert {\bf r}^*-{\bf r}'\vert} \nonumber\\
& & + \frac{\epsilon_1-\epsilon_2}{(\epsilon_1+\epsilon_2)^2}
\int_0^1 {\rm d}x \frac{x^{-\nu}}{\sqrt{a^2+(rr'x)^2/a^2-2rr'x\cos\theta}}
\end{eqnarray}
for $r<a$ and $r'<a$,
\begin{equation}
\beta \langle \phi({\bf r}) \phi({\bf r}') \rangle^{\rm T}
= \left( 1-\frac{2}{\epsilon_1+\epsilon_2} \right) 
\frac{1}{\vert {\bf r}-{\bf r}'\vert} 
+ \frac{\epsilon_1-\epsilon_2}{(\epsilon_1+\epsilon_2)^2}
\int_0^1 {\rm d}x \frac{x^{-\nu}}{\sqrt{r^2+(r'x)^2-2rr'x\cos\theta}}
\end{equation}
for $r>a$ and $r'<a$.
Here, the exponent $\nu=\epsilon_2/(\epsilon_1+\epsilon_2)$ lies in
the interval $[0,1]$.

The radial component of the electric field, normal to the surface of 
the sphere, is given by 
$E_n({\bf r}) = - \partial_r\phi({\bf r})$.
Thus the counterpart of (\ref{field}) reads
$\langle E_n({\bf r}) E_n({\bf r}') \rangle^{\rm T} =
\partial^2_{rr'} \langle \phi({\bf r}) \phi({\bf r}') \rangle^{\rm T}$.
Using the relation (\ref{surfacechargecorr}), with $E_x^{\pm}$
substituted by $E_n^{\pm}$, after lengthy algebra we find the surface 
charge correlation function between two points ${\bf R}$ and ${\bf R}'$ 
on the sphere interface of the form
\begin{equation} \label{result}
\beta \langle \sigma({\bf R}) \sigma({\bf R}') \rangle^{\rm T} 
= - \frac{1}{8\pi^2} \left( \frac{1}{\epsilon_1}
+ \frac{1}{\epsilon_2} - \frac{4}{\epsilon_1+\epsilon_2} \right)
\frac{1}{\vert {\bf R}-{\bf R}'\vert^3} + 
\frac{1}{16\pi^2} \frac{\epsilon_1-\epsilon_2}{(\epsilon_1+\epsilon_2)^2}
\frac{1}{a^3} I(\nu,\cos\theta) ,
\end{equation}
where the integral
\begin{equation} \label{subresult}
I(\nu,\cos\theta) = \int_0^1 {\rm d}x x^{-\nu} \left[
\frac{3(1-x^2)^2}{(1+x^2-2x\cos\theta)^{5/2}} -
\frac{2(1+x^2)}{(1+x^2-2x\cos\theta)^{3/2}} \right] .
\end{equation}
Here, the angle $\theta$ between the points ${\bf R}$ and ${\bf R}'$ 
on the sphere of radius $a$ is given by
$\cos\theta = 1 - \frac{1}{2}(\vert {\bf R}-{\bf R}'\vert/a)^2$.
The above two formulas hold for all macroscopic distances 
$\vert {\bf R}-{\bf R}'\vert$ and sphere radiuses $a$. 

A special case of physical interest is the fixed distance 
$\vert {\bf R}-{\bf R}'\vert$ and the large radius limit 
$\vert {\bf R}-{\bf R}'\vert/a\to 0$.
Writing $1+x^2-2x\cos\theta = (1-x)^2 + \varepsilon x$ with 
$\varepsilon=(\vert {\bf R}-{\bf R}'\vert/a)^2 \ll 1$ and
using the small-$\varepsilon$ expansions
\begin{equation} 
\int_0^1 {\rm d}x \frac{x^{\mu}}{[(1-x)^2+\varepsilon x]^{5/2}}
= \frac{2}{3\varepsilon^2} + \frac{3-2\mu}{6\varepsilon^{3/2}} 
+ \frac{2-3\mu+\mu^2}{6\varepsilon} + \cdots ,
\end{equation}
\begin{equation}
\int_0^1 {\rm d}x \frac{x^{\mu}}{[(1-x)^2+\varepsilon x]^{3/2}}
= \frac{1}{\varepsilon} + \cdots
\end{equation}
($\mu>-1$), we obtain $I(\nu,\cos\theta)=2/\varepsilon+{\cal O}(1)$. 
Thus from (\ref{result}) we conclude that
\begin{eqnarray}
\beta \langle \sigma({\bf R}) \sigma({\bf R}') \rangle^{\rm T} 
& = & - \frac{1}{8\pi^2} \left( \frac{1}{\epsilon_1}
+ \frac{1}{\epsilon_2} - \frac{4}{\epsilon_1+\epsilon_2} \right)
\frac{1}{\vert {\bf R}-{\bf R}'\vert^3} \nonumber \\
& & + \frac{1}{8\pi^2} \frac{\epsilon_1-\epsilon_2}{(\epsilon_1+\epsilon_2)^2}
\frac{1}{\vert {\bf R}-{\bf R}'\vert^2} \frac{1}{a} + 
{\cal O}\left( \frac{1}{a^3} \right) . 
\end{eqnarray}
The leading term, invariant with respect to the media interchange
$\epsilon_1\leftrightarrow\epsilon_2$, corresponds to the classical static 
result for the flat interface (\ref{classstat}). 
The first curvature correction changes the sign under
$\epsilon_1\leftrightarrow\epsilon_2$; this is expected since the
media interchange is effectively equivalent to the change of
the interface curvature to the opposite one.

Another interesting case is the antipode configuration with 
$\vert {\bf R}-{\bf R}' \vert = 2a$, for which curvature effects 
are not erased.
Substituting $\cos\theta = -1$ in the integral (\ref{subresult}), 
$I(\nu,-1)$ is expressible in terms of harmonic numbers \cite{Gradshteyn}.

An open question is a sphere extension of the quantum time-dependent result 
(\ref{quantumr}), derived for a flat interface by taking into account
the retardation.  
This result should be recovered in the large radius limit 
$\vert {\bf R}-{\bf R}'\vert \ll a$, while curvature effects might
invalidate it for $\vert {\bf R}-{\bf R}'\vert \sim a$.
The answer requires the application of Rytov's EM fluctuational theory 
\cite{Rytov} to curved interfaces between media which is a difficult task
left for near future.

\begin{acknowledgement}
The support received from the grants VEGA No. 2/0113/2009 and CE-SAS QUTE
is acknowledged.
\end{acknowledgement}

\end{document}